\documentclass[aip,nobibnotes,nofootinbib,superscriptaddress,twocolumn,
reprint,
]{revtex4-1}

\usepackage{graphicx} 	
\usepackage{dcolumn} 	
\usepackage{bm}   		
\usepackage{amssymb}   	
\usepackage[separate-uncertainty=true,range-units=single,range-phrase=--,retain-explicit-plus=true]{siunitx} 
\usepackage[svgnames]{xcolor}	
\usepackage[version=3]{mhchem}	

\usepackage{xr} 
\usepackage[colorlinks=true]{hyperref}
\usepackage[caption=false]{subfig}
\usepackage[nameinlink,capitalise]{cleveref}
\usepackage{lipsum}

\DeclareSIUnit\unitcell{uc}

\newif\ifsuppinpaper \suppinpapertrue

\ifsuppinpaper
\else
\fi

\def\Vs{V_\mathrm{SQUID}}

\begin{document}
\title{Diamagnetic-like response from localised heating of a paramagnetic material}

\author{Giordano Mattoni}
\email{mattoni@scphys.kyoto-u.ac.jp}
\affiliation{Department of Physics, Graduate School of Science, Kyoto University, Kyoto 606-8502, Japan}

\author{Shingo Yonezawa}
\affiliation{Department of Physics, Graduate School of Science, Kyoto University, Kyoto 606-8502, Japan}

\author{Yoshiteru Maeno}
\affiliation{Department of Physics, Graduate School of Science, Kyoto University, Kyoto 606-8502, Japan}

\begin{abstract}
		In the search of material properties out-of-equilibrium,
	the non-equilibrium steady states induced by electric current are an appealing research direction where unconventional states may emerge.
However, the unavoidable Joule heating caused by flowing current calls for the development of new measurement protocols, with a particular attention to the physical properties of the background materials involved.
		Here, we demonstrate that localised heating can give rise to a large, spurious diamagnetic-like signal.
		This occurs due to the local reduction of the background magnetisation caused by the heated sample, provided that the background material has a Curie-like susceptibility.
	Our experimental results, along with numerical calculations, constitute an important building block for performing accurate magnetic measurements under the flow of electric current.
	
\end{abstract}

\date{\today}
\maketitle
Controlling the physical properties of quantum materials by external parameters constitutes a fundamental basis of modern condensed matter physics \cite{
	mathur1998magnetically,
	zubko2011interface,
	shibauchi2014quantum,
	keimer2015quantum,
	keimer2017physics,
	basov2017towards,
	manca2019large
}.
Recent experiments involving non-equilibrium conditions in strongly correlated compounds sparked interest in the scientific community, with the appeal to uncover unprecedented material properties \cite{
	fausti2011light,
	stojchevska2014ultrafast,
	kagawa2017quenching,
	mattoni2018light
}.
In this regard, non-equilibrium steady states triggered by an electric current are particularly appealing for integrated electronics, with the reported emergence of exotic electronic and magnetic states \cite{
	kumai1999current,
	myers1999current,
	iwasaki2013current,
	okazaki2013current
}.
	Several works by independent research groups recently focused on the non-equilibrium steady states in single-crystal ruthenates \cite{
		zhang2019nano,
		bertinshaw2019unique,
		cirillo2019emergence,
		zhao2019nonequilibrium
	}.
	
	Measuring materials out of thermodynamic equilibrium requires the development of new techniques and measurement protocols \cite{
		nakamura2013electric,
		fursich2019raman
	}.
This is an exceptionally challenging task when performing magnetic measurements at low temperatures because Joule heating becomes a crucial issue.
	For this reason, it is necessary to use additional components that enhance the sample cooling but at the same time provide minimal magnetic background.
	In general, magnetic differential measurements, such as the ones performed with a magnetometer equipped with a vibrating sample or a superconducting quantum interference device (SQUID), cancel the background contribution of spatially homogeneous materials \cite{
		foner1959versatile,
		lewis1996sample
	}.
	However, intense heating due to the Joule effect can locally alter the background material properties, possibly leading to unexpected results.

In this Letter, we report the insurgence of a strong diamagnetic-like signal arising from a paramagnetic material subjected to localised heating.
	We first explicitly demonstrate how diamagnetic-like signals can appear by considering a spatially inhomogeneous mass profile, where a position-dependent magnetisation emerges.
	We then describe our main result: a diamagnetic-like signal can emerge due to localised heating of any background material whose magnetic susceptibility increases with lowering temperature.
We find that these extrinsic effects account for the main part of the diamagnetic signals previously reported in Refs. \cite{
		sow2017current,
		sow2019situ
	}, where their origin was misinterpreted.
The effect presented in the following potentially appears in several techniques of magnetic differential measurement (\cref{fig:OtherDiffTechniques}).
Our results give important guidelines to study magnetic properties in non-equilibrium steady states induced by electric currents.

\begin{figure}[tb]
	\includegraphics[page=1,width=89mm]{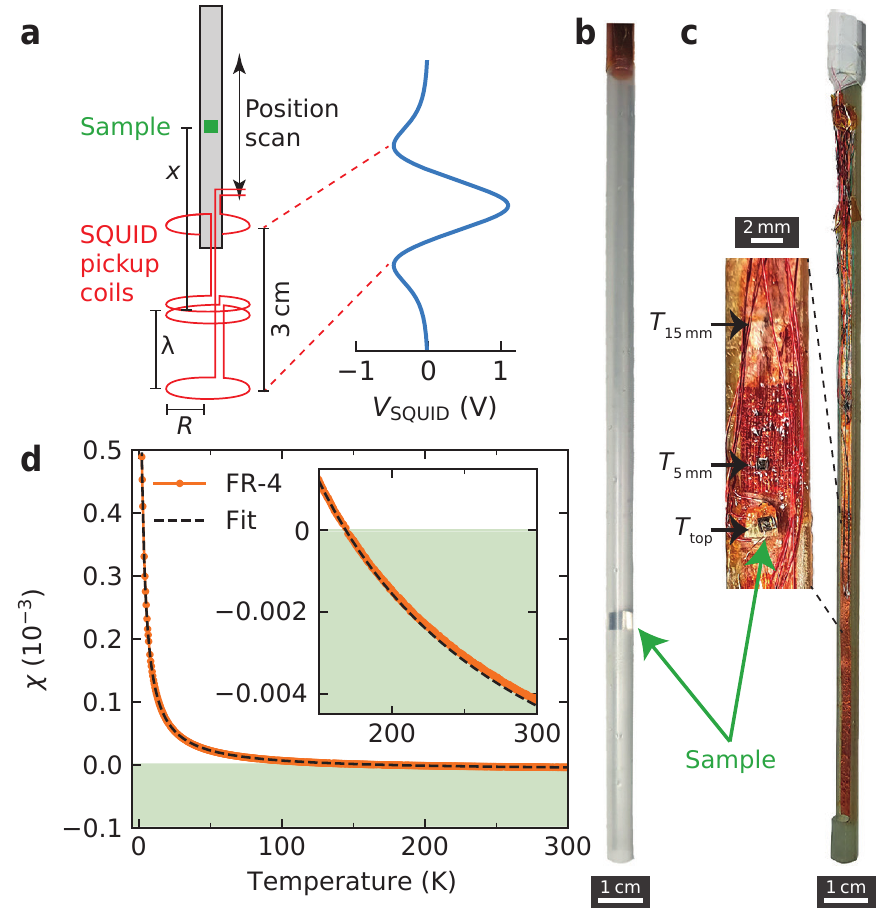}
	\subfloat{\label{fig:SQUID}}
	\subfloat{\label{fig:Design_std}}
	\subfloat{\label{fig:Design_Sow}}
	\subfloat{\label{fig:GlassFiber}}
	\caption{\textbf{Description of the experimental setup.}
			\protect\subref{fig:SQUID} Schematics of the measurement geometry used in a bulk SQUID magnetometer.
			The blue curve shows the typical magnetic signal as a function of sample position.
		\protect\subref{fig:Design_std} Common sample holder made of a plastic straw with a palladium-foil sample, which is mechanically-suspended.
		\protect\subref{fig:Design_Sow} Home-made FR-4 glass epoxy sample holder for measurements with electrical current.
		The close-up photo shows the copper wires glued on the holder surface, the sample, and three temperature sensors placed at different distances.
		\protect\subref{fig:GlassFiber} Temperature dependence of the volume susceptibility of FR-4 at $B_\mathrm{ext}=\SI{1}{\tesla}$.
		The trend has a strong upturn at low temperature and is well-fit by the Curie--Weiss law of \cref{eq:CurieWeiss}.
	}
	\label{fig:Setup}
	
\end{figure}

In the present work, we focus on the specific case of a bulk SQUID magnetometer by using the Magnetic Properties Measurement System (MPMS) by Quantum Design, a common instrument to study magnetic materials.
A SQUID magnetometer measures the magnetisation by scanning the sample up and down through its highly balanced second-derivative coil set (\cref{fig:SQUID}).
When a magnetic sample is moved through such coils, the change of the enclosed magnetic flux is detected by the coupled SQUID as a voltage.
Assuming a point-like magnetic sample whose position $x$ is scanned through the coils,
the signal is given by
\cite{
	lewis1996sample
}
\begin{widetext}
	\begin{equation}
	\label{eq:SQUID}
	V_\mathrm{SQUID,\,pt}(x) = C_\mathrm{cal}C_\mathrm{range}\mu_\mathrm{tot} \left[{
		\frac{2}{\sqrt{[R^2+x^2]^3}}
		- \frac{1}{\sqrt{[R^2+(x - \lambda)^2]^3}}
		- \frac{1}{\sqrt{[R^2+(x + \lambda)^2]^3}}
		+ mx + q
	}\right],
	\end{equation}
\end{widetext}
with
instrument calibration constant $C_\mathrm{cal}$,
selected magnetisation range $C_\mathrm{range}$,
total sample magnetisation $\mu_\mathrm{tot}$,
radius of the coils $R$,
their distance $\lambda$, and
a possible voltage drift $mx + q$
(values for our SQUID magnetometer are given in the Supplementary Material).
The sample is measured in a low-pressure helium environment (about \SI{0.1}{\milli\bar}), and its temperature is controlled by heating/cooling the walls of the sample space
that are controlled by a built-in thermometer.
We use this thermometer to readout the refigerator temperature that is presented in the following.

Since it is important to minimise background magnetic signals, the most-common sample holder for SQUID magnetometers consists of a thin plastic straw where the sample is mechanically fixed, if possible without the use of any glue (\cref{fig:Design_std}).
The straw is much longer than the \SI{3}{\centi\metre} distance between the coils, so that it provides an ideal homogeneous background.
However, for measurements with a current flowing in the sample this simple design cannot be used for three main reasons:
(1) wires are necessary to provide electrical contact to the sample,
(2) it is necessary to carry away the heat generated in the sample, and
(3) a thermometer is required to accurately read the sample temperature.
For these reasons, it is necessary to introduce an additional material that works as a sample holder.
In \cref{fig:Design_Sow} we show the sample holder that was previously designed for the experiments in Refs. \cite{
	sow2017current,
	sow2019situ
}.
The holder is made of the glass epoxy FR-4, a readily available material which is often used for printed circuit boards (PCBs).
We chose this material because it has a good mechanical stability, necessary to support the sample while it is moved through the coil set, and it is rather simple to machine accurately,
thus allowing to build a sample holder with a well-defined geometry.
A strip of insulated copper wires is glued on the FR-4 surface in order to homogenize the temperature distribution, while minimising heating effects due to possible Eddy currents.

We study the magnetic properties of the FR-4 material by cutting a small piece (\SI{53}{\milli\gram}) and measuring it in the straw sample holder.
We present in \cref{fig:GlassFiber} the temperature dependence of the magnetic susceptibility calculated as
$\chi(T) = \frac{\mu_0\rho}{mB_\mathrm{ext}}\mu_\mathrm{tot}(T)$,
with vacuum permeability $\mu_0$,
sample density $\rho$,
mass $m$,
and applied magnetic field $B_\mathrm{ext}$.
For the whole data presented in the main text, we fix $B_\mathrm{ext}=\SI{1}{\tesla}$ (additional characterisation for different values of $B_\mathrm{ext}$ in \cref{fig:EpoxyMvsB}).
As expected for a plastic, the material is weakly diamagnetic at room temperature with $\chi\sim\SI{-4e-6}{}$ (for reference, $\chi = 1\ \mathrm{[SI]} = \frac{1}{4\pi}\,\mathrm{emu/cm}^3\ \mathrm{[cgs]}$).
Upon cooling, the material becomes paramagnetic below about \SI{170}{\kelvin}, and shows a strong upturn at low temperature.
We fit this experimental trend with the Curie--Weiss law
\begin{equation}
\label{eq:CurieWeiss}
\chi(T) = \frac{C}{T-T_\mathrm{CW}} + \chi_\infty.
\end{equation}
The fit (black dashed line in \cref{fig:GlassFiber}) shows good agreement with the experimental data.
We find as optimal parameters
$C = \SI{1.67e-3}{\kelvin}$,
$T_\mathrm{CW}=\SI{-1.2}{\kelvin}$,
and $\chi_\infty=\SI{-9.8e-6}{}$, with the value of $C$ comparable to the one previously reported for glass epoxy \cite{
	lockhart1990magnetic
}.
The increasing trend of $\chi(T)$ for lowering temperatures indicates that heating can decrease the FR-4 magnetisation, at any temperature.
We investigate in the following what are the implications of local temperature changes.

\begin{figure}[tb]
	\includegraphics[page=2,width=89mm]{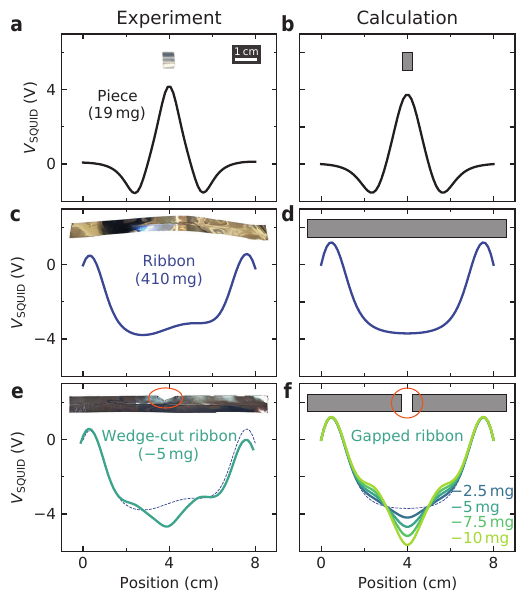}
	
	\subfloat{\label{fig:Pd_sq_exp}}
	\subfloat{\label{fig:Pd_sq_sim}}
	\subfloat{\label{fig:Pd_rib_exp}}
	\subfloat{\label{fig:Pd_rib_sim}}
	\subfloat{\label{fig:Pd_cut_exp}}
	\subfloat{\label{fig:Pd_cut_sim}}
	\caption{\textbf{Magnetic signal of an inhomogeneous palladium ribbon.}
		Experimental measurements at $T=\SI{300}{\kelvin}$, $B_\mathrm{ext}=\SI{1}{\tesla}$ of a Pd foil (thickness \SI{50}{\micro\metre}, width \SI{0.8}{\centi\metre}) in the shape of:
		\protect\subref{fig:Pd_sq_exp} a \SI{0.4}{\centi\metre} piece,
		\protect\subref{fig:Pd_rib_exp} a \SI{9}{\centi\metre} ribbon, 
		\protect\subref{fig:Pd_cut_exp} and the same ribbon with a wedge cut.
		\protect\subref{fig:Pd_sq_sim}, \protect\subref{fig:Pd_rib_sim}, and \protect\subref{fig:Pd_cut_sim} are a numerical calculation of the data in the left column obtained with \cref{eq:SQUID_int}.
	}
	\label{fig:Pd_simulation}
	
\end{figure}

To this goal, we first explicitly show
the magnetic signal of a material with a controlled inhomogeneity by using a palladium (Pd) foil as reference.
Pieces of this foil are measured at room temperature ($T=\SI{300}{\kelvin}$) in the straw sample holder of \cref{fig:Design_std}.
We show in \cref{fig:Pd_sq_exp} the raw SQUID signal of a \SI{0.4}{\centi\metre} Pd piece, which shows a paramagnetic response.
It is possible to numerically calculate the signal of a magnetic sample, with a position-dependent magnetic moment $\mu(x')$, by computing over the material length $L$ the convolution integral
\begin{equation}
\label{eq:SQUID_int}
\Vs(x) =
\int_{L}\frac{\mu(x')}{\mu_\mathrm{tot}}V_\mathrm{SQUID,\,pt}(x-x') dx'.
\end{equation}
Using this equation, we calculate in \cref{fig:Pd_sq_sim} the signal expected from the Pd piece, finding a result compatible with the experiment.
In the calculation, the Pd foil is treated as a monodimensional segment, a reasonable approximation due to the symmetry of this experimental configuration.
In \cref{fig:Pd_rib_exp}, we show the SQUID signal of a homogeneous Pd ribbon of length \SI{9}{\centi\metre}.
The signal shows a rather flat region of about \SI{4}{\centi\metre} around the centre and two peaks \SI{7}{\centi\metre} apart.
We calculate in \cref{fig:Pd_rib_sim} the signal of the Pd ribbon by using \cref{eq:SQUID_int}, and find again good agreement with the experimental data in \cref{fig:Pd_rib_exp}, apart from some small deviations probably related to imperfections of the Pd foil.

To display the magnetic signal of a controlled inhomogeneity, we now remove a small part of the Pd ribbon by a wedge cut.
The cut is about \SI{0.5}{\centi\metre} in length and it allows to locally reduce the mass, while mechanically keeping the material as one piece.
As shown in \cref{fig:Pd_cut_exp}, the wedge cut produces a dip in $\Vs$ at about $x=\SI{4}{\centi\metre}$, while the rest of the signal remains pretty much unaltered (blue dashed line for reference).
The observed dip resembles the signal of a diamagnetic material placed at $x=\SI{4}{\centi\metre}$, even though there is no diamagnetic material here (Pd is paramagnetic).
This occurs because the SQUID magnetometer detects the contrast of magnetisation between one point and its surroundings.
We calculate this effect with \cref{eq:SQUID_int} by considering a Pd ribbon with a small gap in the middle.
Even if this configuration is slightly different from the experimental one, it remains simple enough to demonstrate the nature of the effect.
The signals calculated for different gap sizes are shown in \cref{fig:Pd_cut_sim}, where the values of the removed masses are indicated.
When the mass of the gap corresponds to the one of the wedge cut (\SI{-5}{\milli\gram}), the experimental and the calculated data resemble each other,
confirming that the diamagnetic-like dip can be generated by the controlled inhomogeneity in the paramagnetic material.

\begin{figure}[tb]
	\includegraphics[page=3,width=89mm]{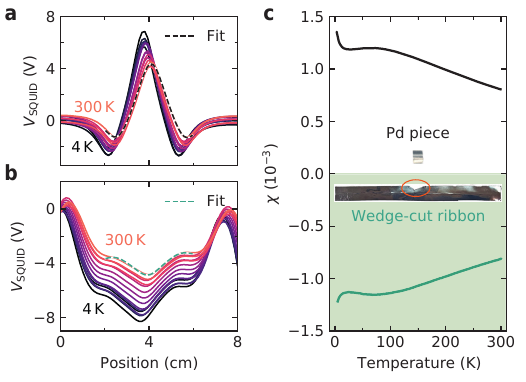}
	
	\subfloat{\label{fig:SQUID_square}}
	\subfloat{\label{fig:SQUID_wedge}}
	\subfloat{\label{fig:MvsT_Chi}}
	\caption{\textbf{Temperature dependence of the magnetic signal of two Pd samples.}
		\protect\subref{fig:SQUID_square} SQUID voltage for the Pd piece and
		\protect\subref{fig:SQUID_wedge} the wedge-cut Pd ribbon as a function of temperature at $B_\mathrm{ext}=\SI{1}{\tesla}$.
		\protect\subref{fig:MvsT_Chi} Volume susceptibility obtained by fitting the SQUID signals in panels \protect\subref{fig:SQUID_square} and \protect\subref{fig:SQUID_wedge}.
	}
	\label{fig:MvsT}
	
\end{figure}

\begin{figure*}[!tb]
	\includegraphics[page=4,width=155mm]{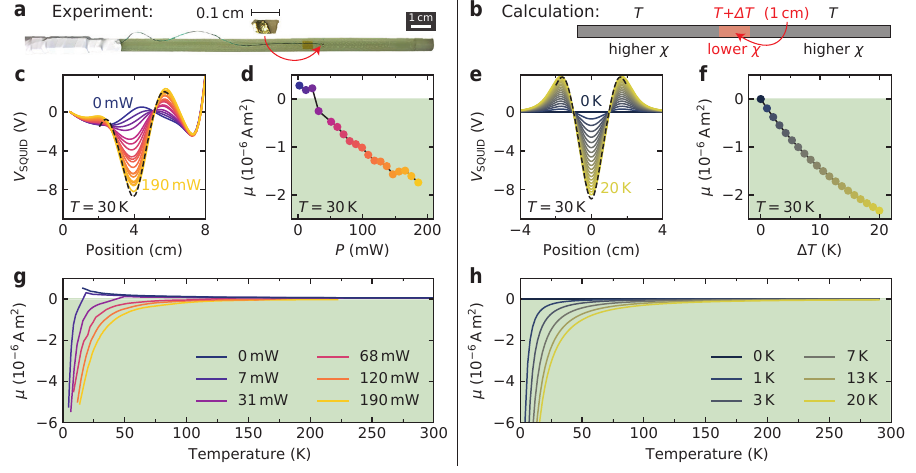}
	
	\subfloat{\label{fig:Heating_exp_img}}
	\subfloat{\label{fig:Heating_sim_img}}
	\subfloat{\label{fig:Heating_exp_Pos}}
	\subfloat{\label{fig:Heating_exp_P}}
	\subfloat{\label{fig:Heating_sim_Pos}}
	\subfloat{\label{fig:Heating_sim_dT}}
	\subfloat{\label{fig:Heating_exp_T}}
	\subfloat{\label{fig:Heating_sim_T}}
	\caption{\textbf{Diamagnetic-like response from localised heating of the FR-4 material.}
		\protect\subref{fig:Heating_exp_img} Photograph of the FR-4 sample holder with a \SI{1.8}{\kilo\ohm} \ce{RuO2} resistor to which a continuous electrical power $P$ is supplied.
		\protect\subref{fig:Heating_sim_img} Schematic of the material used for the numerical calculation, where a region of \SI{1}{\centi\metre} is heated up by $\Delta T$.
		\protect\subref{fig:Heating_exp_Pos} SQUID signal at $T=\SI{30}{\kelvin}$ and $B_\mathrm{ext}=\SI{1}{\tesla}$
		for different values of $P$ and
		\protect\subref{fig:Heating_exp_P} ``magnetic moment" obtained by a \SI{4}{\centi\metre} fit (dashed black line) of $\Vs$.
		\protect\subref{fig:Heating_sim_Pos} and 
		\protect\subref{fig:Heating_sim_dT} are the corresponding numerical calculations of the locally-heated FR-4 material for different values of $\Delta T$.
		\protect\subref{fig:Heating_exp_T} Experimental and
		\protect\subref{fig:Heating_sim_T} calculated temperature dependence of the magnetic signal caused by localised heating.
	}
	\label{fig:HeatingSimulation}
	
\end{figure*}

We now study the temperature dependence of the magnetic signal of the Pd samples of \cref{fig:Pd_sq_exp,fig:Pd_cut_exp}.
We extract $\mu$ by fitting with \cref{eq:SQUID} a portion of \SI{4}{\centi\metre} of $\Vs$ around the sample position at different temperatures.
The peak-like shape in \cref{fig:SQUID_square} increases in magnitude upon cooling, indicating an increase in paramagnetic signal.
The small shift of the peak position to the left is due to the thermal shrinkage of the holder components upon cooling.
We extract the corresponding volume susceptibility $\chi(T)$ in \cref{fig:MvsT_Chi}, finding good agreement with previous reports on Pd \cite{
	kriessman1954magnetic
}.
The dip-like shape produced by the wedge-cut ribbon in \cref{fig:SQUID_wedge}, instead, is well-fit as a diamagnetic signal.
We extract the corresponding $\chi(T)$ in \cref{fig:MvsT_Chi} by using the mass removed from the wedge cut as the ``sample mass".
The resulting curve has essentially the same shape and magnitude of the Pd piece, but with opposite sign.
This means that the magnetisation of the hole left by the removed part of a paramagnetic material behaves as its opposite, that is like a diamagnetic material.
This observation can be generalised to state that a magnetic inhomogeneity in a certain material, which locally reduces its magnetisation, produces a diamagnetic-like signal.

We now move on to the main part of this Letter to show how a diamagnetic-like signal can be produced by localised heating of a background material.
We demonstrate the effect by using as a test background material the FR-4 sample holder that was previously developed in Refs. \cite{
	sow2017current,
	sow2019situ
}
(the signal of the holder alone is shown in \cref{fig:Holder}).
We use a commercial \ce{RuO2} resistor (surface mount device, size \SI{0.5x1}{\milli\metre}) to provide a local electrical power $P$ by means of a continuous current (\cref{fig:Heating_exp_img}).
To make the magnetic signal from the resistor mostly negligible in the performed measurements, we mechanically remove the original contact pads and use Ag epoxy to provide electrical contact (\cref{fig:Resistor}).
This treatment reduced the magnetisation of the resistor by more than 50 times.
As shown in \cref{fig:Heating_exp_Pos}, at $T=\SI{30}{\kelvin}$ the electric power supplied to the resistor progressively changes $\Vs$ from a peak (paramagnetic-like) to a dip (diamagnetic-like).
The maximum power used here is comparable to what used in previous experiments on current-induced non-equilibrium steady states, where it was shown that the resulting heating strongly depends on the sample cooling conditions \cite{
	fursich2019raman
}.
By fitting the curves, we extract the corresponding ``magnetic moment" $\mu$ in \cref{fig:Heating_exp_P}, which becomes more negative with increasing power (for reference, $\mu = \SI{e-3}{\ampere\square\metre}\ \mathrm{[SI]} = 1\,\mathrm{emu}\  \mathrm{[cgs]}$).
We show in \cref{fig:HolderCurrent_YesNo} that this change is localised around the position of the resistor and it is independent of the sign of the supplied current.

To better understand the nature of this observed effect, we take into account a simple model of a spatially homogeneous FR-4 material at a temperature $T=\SI{30}{\kelvin}$.
We consider a central region to be heated to a temperature $T+ \Delta T$, decreasing its magnetisation.
Even though actual distributions of temperature may follow a profile smoother than this abrupt step function, this model is simple enough to demonstrate the effects of heating.
By using \cref{eq:SQUID_int} and the $\chi(T)$ from \cref{fig:GlassFiber}, we numerically calculate $\Vs$ (\cref{fig:Heating_sim_Pos}) and the corresponding ``magnetic moment" (\cref{fig:Heating_sim_dT}) for different values of $\Delta T$.
The results of this calculation are comparable to the experimental data, showing an increasing diamagnetic-like signal for larger values of $\Delta T$.
We show in \cref{fig:Heating_Extended} that the experimental and calculated data are in analogous agreement also for other selected temperatures.

Finally, we experimentally measure in \cref{fig:Heating_exp_T} the temperature dependence of the ``magnetic moment" generated by different values of electric power through the resistor.
Above \SI{100}{\kelvin}, the curves are flat, consistent with the rather flat temperature dependence of the FR-4 $\chi (T)$ (\cref{fig:GlassFiber}).
At lower temperatures, a strong diamagnetic-like response is measured even at the smallest applied power.
We note that the curves here are limited by the cooling power of our measurement system, so that it is not possible to reach the base temperature of \SI{2}{\kelvin} in the presence of large Joule heating.
We numerically calculate the temperature dependence of the ``magnetic moment" of the locally heated FR-4 in \cref{fig:Heating_sim_T}, and find that even for a $\Delta T$ as small as \SI{1}{\kelvin}, a large diamagnetic-like signal appears at low temperatures.
The data is in remarkable agreement with the experimental curves, supporting the interpretation that the observed diamagnetic-like response is caused by localised heating of the background material.
On leaving this paragraph, we should emphasise that the results presented here are not specific to FR-4, but apply to most materials, provided that their $\chi (T)$ increases at low temperatures \cite{
	lockhart1990magnetic
}.
If the studied magnetic signal is very small, even common background materials such as fused silica, copper, or other plastics can give rise to spurious diamagnetic-like signals.

An important consequence of this work is the reinterpretation of the strong diamagnetic-like signals reported in Refs. \cite{
	sow2017current,
	sow2019situ
}
to be due to a spurious origin.
As we show in the preliminary data of \cref{fig:CRO_Diamagnetism}, where only a poor sample cooling is provided, the magnitude of \ce{Ca2RuO4} magnetisation is only weakly affected by applied current when a sample holder with small magnetic signal is used.

To conclude, low-temperature magnetic measurements of highly resistive materials in current-induced non-equilibrium steady states require particular care in the experimental design.
This applies to bulk SQUID magnetometry and also to other techniques for differential magnetic measurements.
It is crucial to carefully regulate the heat dissipation, to achieve proper sample cooling, and at the same time to avoid localised heating.
If these conditions are not met, localised heating of background materials can create spurious signals having magnitude comparable or even greater than the sample under consideration.
This work is an important step towards obtaining accurate magnetic measurements of materials under applied electrical current.

\section*{Acknowledgements}
The authors thank N. Manca for providing valuable feedback on the manuscript.
This work was supported by JSPS Grant-in-Aids KAKENHI Nos. JP26247060, JP15H05852, JP15K21717, JP17H06136, and JP18K04715, as well as by JSPS Core-to-Core program.
G.M. acknowledges support from the Dutch Research Council (NWO) through a Rubicon grant number 019.183EN.031.

\section*{Data availability statement}
The data that supports the findings of this study is available from the corresponding author
upon reasonable request.

\section*{Supplementary material}
See the supplementary material for the parameters of our SQUID magnetometer, other possible techniques of magnetic differential measurements, additional experimental and calculated data on the FR-4 material, reinterpretation of previous data on \ce{Ca2RuO4}.

\bibliography{Biblio_BG_Dianagnetism}

\ifsuppinpaper

\ifsuppinpaper

	\onecolumngrid
	\appendix
	\newpage
	\noindent\rule{1\columnwidth}{1pt}
	\section*{\huge \texttt{Supplementary Material}}
	\noindent\rule{1\columnwidth}{1pt}

\else

\documentclass[
notitlepage
]{revtex4-2}

\usepackage{graphicx} 	
\usepackage{dcolumn} 	
\usepackage{bm}   		
\usepackage{amssymb}   	
\usepackage[separate-uncertainty=true,range-units=single,range-phrase=--,retain-explicit-plus=true]{siunitx} 
\usepackage[svgnames]{xcolor}	
\usepackage[version=3]{mhchem}	

\hyphenation{ALPGEN}
\hyphenation{EVTGEN}
\hyphenation{PYTHIA}

\usepackage{xr} 
\usepackage[hidelinks]{hyperref}
\usepackage[caption=false]{subfig}
\usepackage[nameinlink,capitalise]{cleveref}
\usepackage{lipsum}

\DeclareSIUnit\unitcell{uc}

	\externaldocument{BG_Diamagnetism}
	\begin{document}
	\title{\texttt{Supplementary Material}\\
		Diamagnetic-like response arising from localised heating of a paramagnetic material
	}

\author{Giordano Mattoni}\email{mattoni@scphys.kyoto-u.ac.jp}
\affiliation{Department of Physics, Graduate School of Science, Kyoto University, Kyoto 606-8502, Japan}

\author{Shingo Yonezawa}
\affiliation{Department of Physics, Graduate School of Science, Kyoto University, Kyoto 606-8502, Japan}

\author{Yoshiteru Maeno}
\affiliation{Department of Physics, Graduate School of Science, Kyoto University, Kyoto 606-8502, Japan}

	\maketitle
	
\fi

\renewcommand\thefigure{S\arabic{figure}}    
\setcounter{figure}{0}
\renewcommand\thetable{S\arabic{table}}    
\setcounter{table}{0}
\renewcommand\theequation{S\arabic{equation}}    
\setcounter{equation}{0}

\vfill
\section{Parameters used in EQ. (1) for our SQUID magnetometer (Quantum Design MPMS XL).}

\begin{eqnarray}
&C_\mathrm{cal} = \SI{0.198}{\per\volt}\\
&C_\mathrm{range} = \SI{0.025e-3}{\ampere \square \metre}\\
&R=\SI{0.97}{\centi\metre}\\
&\lambda=\SI{1.519}{\centi\metre}
\end{eqnarray}
\vfill

\section{Supplementary figures.}

\begin{figure*}[h]
	\includegraphics[page=5,width=180mm]{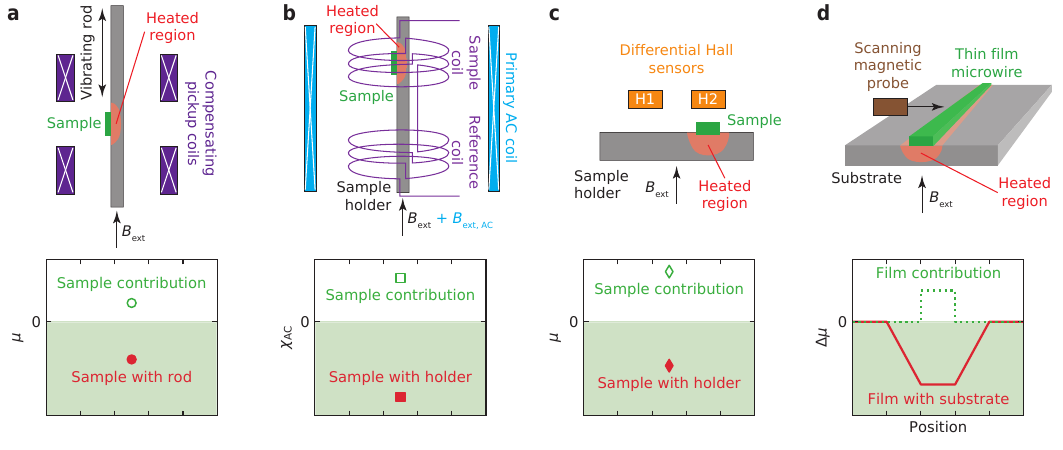}
	
	\subfloat{\label{fig:OtherDiffTechniques_VSM}}
	\subfloat{\label{fig:OtherDiffTechniques_AC}}
	\subfloat{\label{fig:OtherDiffTechniques_Hall}}
	\subfloat{\label{fig:OtherDiffTechniques_Scanning}}	
	
	\caption{
		\textbf{Possible techniques of differential magnetic measurements where spurious diamagnetic-like signals can arise with applied current.}
		In the presented cases, a current flowing in the sample produces localised heating  of the background materials (red shaded area) due to the Joule effect.
		As discussed in the main text, a large diamagnetic-like response can arise in these configurations provided that the magnetic susceptibility of the background material increases at low temperature.
		Bulk measurements such as the ones performed with
		\protect\subref{fig:OtherDiffTechniques_VSM} a vibrating sample magnetometer (VSM) or
		\protect\subref{fig:OtherDiffTechniques_AC} AC susceptibility should be certainly affected by the effect discussed in this Letter.
		\protect\subref{fig:OtherDiffTechniques_Hall} Measurements by means of a differential Hall probe, where the magnetisation is evaluated from the difference of the readings of two Hall sensors, can also present this effect.
		\protect\subref{fig:OtherDiffTechniques_Scanning} Moreover, we infer that a similar effect can happen also in microscopic measurements, such as for the case of a thin-film material over a substrate measured by means of a scanning magnetic-probe technique (i.e., SQUID, Hall, or magnetic force microscopy).
		In all cases, a homogeneous magnetic field $B_\mathrm{ext}$ is present in the direction indicated by the arrow.
		Bottom row: Sketch of possible magnetic signals arising from the experimental configurations described above.
		In addition to the presented measurements where an electric current is flowing through the sample, similar effects may occur with other sources of local heating, such as, for example, light irradiation.
	}
	
	\label{fig:OtherDiffTechniques}
	
\end{figure*}

\begin{figure*}[h]
	\includegraphics[page=6,width=110mm]{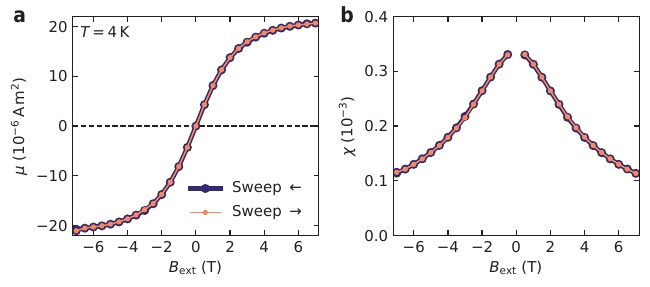}

	\subfloat{\label{fig:EpoxyMvsB_mu}}	
	\subfloat{\label{fig:EpoxyMvsB_chi}}
	
	\caption{\textbf{Magnetic properties of FR-4 glass epoxy as a function of magnetic field.}
		\protect\subref{fig:EpoxyMvsB_mu} Magnetic moment of a \SI{53}{\milli\gram} piece of glass epoxy and
		\protect\subref{fig:EpoxyMvsB_chi} corresponding volume susceptibility both measured at $T=\SI{4}{\kelvin}$.
		The curves saturate with increasing magnetic field, as expected from a paramagnetic material at low temperatures.
		Data for the high-to-low (blue) and low-to-high (orange) magnetic field sweeps overlap, indicating the absence of remnant magnetisation.
	}
	
	\label{fig:EpoxyMvsB}
	
\end{figure*}

\begin{figure*}[h]
	\includegraphics[page=7,width=160mm]{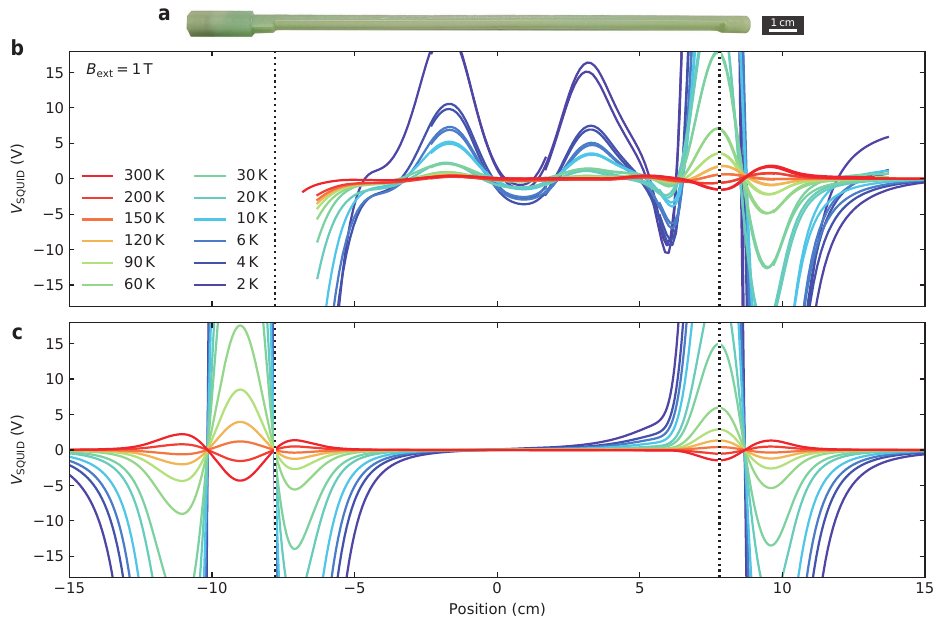}

	\subfloat{\label{fig:Holder_photo}}	
	\subfloat{\label{fig:Holder_exp}}
	\subfloat{\label{fig:Holder_sim}}
	
	\caption{\textbf{Magnetic signal of the home-made sample holder for different temperatures.}
		\protect\subref{fig:Holder_photo} Photograph of the FR-4 glass epoxy sample holder. The photo is aligned with the $x$-axis of the graphs right below.
		\protect\subref{fig:Holder_exp} Experimental SQUID response at different temperatures.
		The vertical dotted lines indicate the edges of the sample-holder part with constant cross-section, designed to host the sample.
		The broad scans are obtained by stitching together several scans of \SI{8}{\centi\metre} in length and are limited on the left side by the end of stroke of the measurement system.
		\protect\subref{fig:Holder_sim} Numerical calculation of the signal based on the real sample holder geometry and a homogeneous FR-4 material.
		The data correctly reproduces the temperature dependence of the experimental signal relative to the edges of the sample holder (dotted lines).
		Two additional peaks centred at \SI{-1.6}{\centi\metre} and \SI{3.3}{\centi\metre} appear only in the experimental data below about \SI{100}{\kelvin}.
		These peaks are most-likely due to spatial inhomogeneity of the FR-4 material used to build the sample holder.
	}
	
	\label{fig:Holder}
	
\end{figure*}

\begin{figure*}[h]
	\includegraphics[page=8,width=100mm]{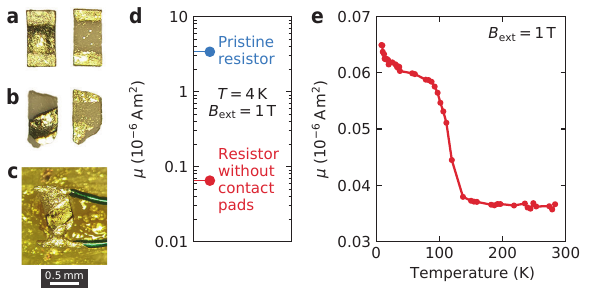}
	
	\subfloat{\label{fig:Resistor_Pristine}}	
	\subfloat{\label{fig:Resistor_Broken}}
	\subfloat{\label{fig:Resistor_Mounted}}
	\subfloat{\label{fig:Resistor_muComparison}}
	\subfloat{\label{fig:Resistor_muvsT}}
	
	\caption{\textbf{Preparation of the \ce{RuO2} resistor with extremely low magnetic signal.}
		\protect\subref{fig:Resistor_Pristine} Photo (front and back) of a pristine \SI{0.5x1}{\milli\metre} surface-mount thick-film \ce{RuO2} resistor of \SI{1.8}{\kilo\ohm} (KOA, RK73K1EJ).
		\protect\subref{fig:Resistor_Broken} Resistor with the contact pads mechanically removed and
		\protect\subref{fig:Resistor_Mounted} mounted on the glass epoxy sample holder with GE varnish.
		Ag epoxy is used to provide electrical contact with two copper wires of diameter \SI{0.12}{\milli\metre}.
		The scale bar applies to all the photos.
		\protect\subref{fig:Resistor_muComparison} The magnetic signal of the resistor (blue) is greatly reduced after the removal of the contact pads (red).
		The measurements are performed at $T=\SI{4}{\kelvin}$ and $B_\mathrm{ext}=\SI{1}{\tesla}$.
		\protect\subref{fig:Resistor_muvsT} The magnetic signal from the \ce{RuO2} resistor after removal of the contact pads is extremely low in the whole explored temperature range, accounting for less than \SI{1}{\percent} of the magnetic signals reported in \cref{fig:HeatingSimulation} of the main text.
	}
	
	\label{fig:Resistor}
	
\end{figure*}

\begin{figure*}[h]
	\includegraphics[page=9,width=130mm]{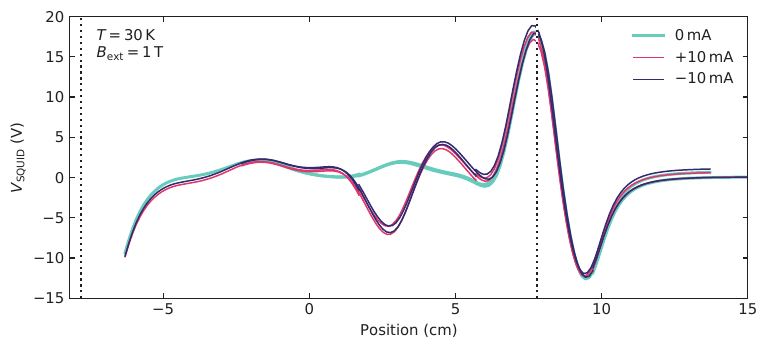}
	
	\caption{\textbf{Broad-range SQUID signal due to localised heating.}
		SQUID signal of the glass epoxy sample holder at $T=\SI{30}{\kelvin}$, $B_\mathrm{ext}=\SI{1}{\tesla}$.
		When a current of \SI{10}{\milli\ampere} (producing \SI{190}{\milli\watt}) is supplied to a \ce{RuO2} resistor positioned at about $x=\SI{3}{\centi\metre}$, a radical change of the signal is observed.
		The positive-to-negative change in SQUID response determines the diamagnetic-like signal, arising from local heating of the sample holder.
		The response is independent of the sign of the current, as expected from an effect dictated by Joule heating.
		The vertical dotted lines indicate the edges of the sample-holder as in \cref{fig:Holder}.
	}
	
	\label{fig:HolderCurrent_YesNo}
	
\end{figure*}

\begin{figure*}[h]
	\includegraphics[page=10,width=150mm]{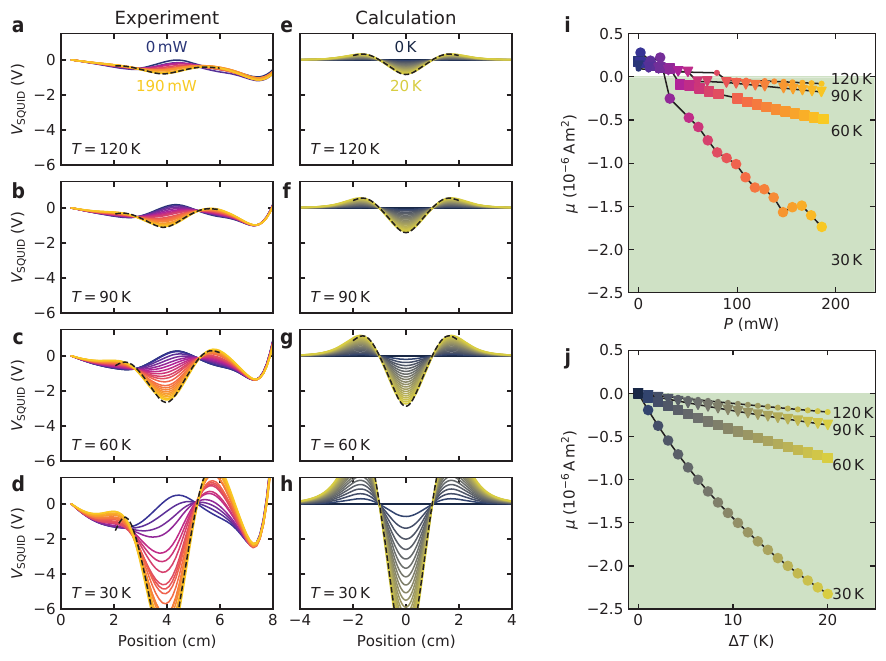}
	
	
	\subfloat{\label{fig:Heating_Extended_a}}
	\subfloat{\label{fig:Heating_Extended_b}}
	\subfloat{\label{fig:Heating_Extended_c}}
	\subfloat{\label{fig:Heating_Extended_d}}
	\subfloat{\label{fig:Heating_Extended_e}}
	\subfloat{\label{fig:Heating_Extended_f}}
	\subfloat{\label{fig:Heating_Extended_g}}
	\subfloat{\label{fig:Heating_Extended_h}}
	\subfloat{\label{fig:Heating_Extended_i}}
	\subfloat{\label{fig:Heating_Extended_j}}
	
	\caption{
		\textbf{Localised heating effects at different temperatures.}
		This figure is an extension of the data presented in \cref{fig:HeatingSimulation} of the main text.
		\protect\subref{fig:Heating_Extended_a} to \protect\subref{fig:Heating_Extended_d} experimental signal of the FR-4 holder locally heated with a resistor.
		At all the selected temperatures, the diamagnetic-like signal becomes stronger with increasing electric power from 0 to \SI{190}{\milli\watt}.
		\protect\subref{fig:Heating_Extended_e} to \protect\subref{fig:Heating_Extended_h} numerical calculations for the locally-heated FR-4 material as a function of $\Delta T$ from 0 to \SI{20}{\kelvin} and different selected base temperatures.
		\protect\subref{fig:Heating_Extended_i} and \protect\subref{fig:Heating_Extended_j}	show the ``magnetic moment" extracted from the data in the left columns.
		The striking similarity between the experimental and calculated data supports localised heating to be the origin of the diamagnetic-like signal.
	}
	
	\label{fig:Heating_Extended}
	
\end{figure*}

\begin{figure*}[h]
	\includegraphics[page=11,width=150mm]{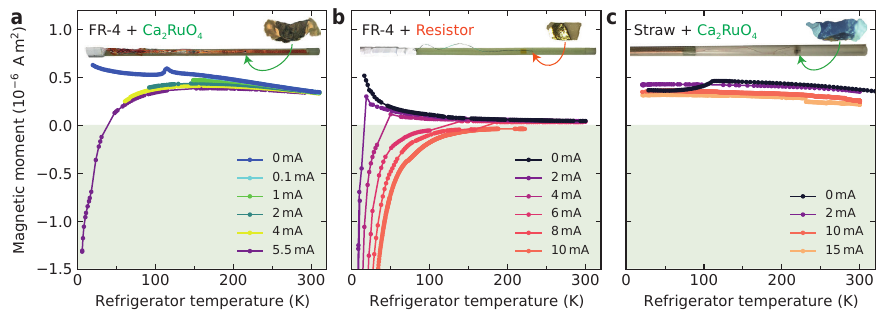}
	
	
	\subfloat{\label{fig:CRO_Diamagnetism_a}}
	\subfloat{\label{fig:CRO_Diamagnetism_b}}
	\subfloat{\label{fig:CRO_Diamagnetism_c}}
	
	\caption{
	\textbf{Implications of the present findings on the current-induced magnetic signals of \ce{Ca2RuO4}.}
	\protect\subref{fig:CRO_Diamagnetism_a} Magnetic moment of \ce{Ca2RuO4} measured with the FR-4 sample holder with the copper strip.
	A strong diamagnetic-like signal appears at low temperature with applied current.
	This data is comparable to the one published in [Sow et al., Science 358 1084 (2017)], with the difference that here the temperature is measured with the refrigerator thermometer.
	\protect\subref{fig:CRO_Diamagnetism_b} Diamagnetic-like response from the locally heated FR-4 presented in \cref{fig:HeatingSimulation} of the main text.
	\protect\subref{fig:CRO_Diamagnetism_c} \ce{Ca2RuO4} measured with applied current by using the straw sample holder.
	In all panels, the actual sample temperature may differ significantly from the refrigerator temperature.
	In panel \protect\subref{fig:CRO_Diamagnetism_c}, this is more significant since the straw sample holder without copper provides poorer sample cooling then the holder used in \protect\subref{fig:CRO_Diamagnetism_a}.
	Hence, the measurements of panel \protect\subref{fig:CRO_Diamagnetism_c} have only preliminary character.
	Nonetheless, they show that the diamagnetic-like signal measured in \protect\subref{fig:CRO_Diamagnetism_a} mostly disappears upon changing the sample holder.
	}
	
	\label{fig:CRO_Diamagnetism}
	
\end{figure*}

\else
\fi

\end{document}